\documentclass[prb,aps,showpacs,floats,twocolumn]{revtex4}%
\usepackage{amssymb}
\usepackage{graphicx}
\usepackage{dcolumn}
\usepackage{bm}
\usepackage{amsmath}
\usepackage{amsfonts}%

\newcommand{\mb}[1]     {\mbox{\boldmath $#1$}}
\newcommand{\bq}{\begin{equation}}
\newcommand{\eq}{\end{equation}}
\newcommand{\bqa}{\begin{eqnarray}}
\newcommand{\eqa}{\end{eqnarray}}
\newcommand{\nn}{\nonumber \\}
\def\be     {\begin{equation}}
\def\ee     {\end{equation}}
\def\bea        {\begin{eqnarray}}
\def\eea        {\end{eqnarray}}
\def\bnn    {\begin{eqnarray*}}
\def\enn    {\end{eqnarray*}}

\begin{document}
\title{A phase diagram for a topological Kondo insulating system}
\author{Minh-Tien Tran$^{1,2}$, Tetsuya Takimoto$^{1,3}$, and Ki-Seok Kim$^{1,3}$}
\affiliation{$^1$Asia Pacific Center for Theoretical Physics,
POSTECH, Pohang, Gyeongbuk 790-784, Republic of Korea \\
$^2$Institute of Physics, Vietnamese Academy of Science and
Technology, P.O.Box 429, 10000 Hanoi, Vietnam \\ $^{3}$Department
of Physics, POSTECH, Pohang, Gyeongbuk 790-784, Republic of Korea}
\date{\today}

\begin{abstract}
The discovery of topological insulators in non-interacting
electron systems has motivated the community to search such
topological states of matter in correlated electrons both
theoretically and experimentally. In this paper we investigate a
phase diagram for a topological Kondo insulating system, where an
emergent ``spin"-dependent Kondo effect gives rise to an inversion
for heavy-fermion bands, responsible for a topological Kondo
insulator. Resorting to the U(1) slave-boson mean-field analysis,
we uncover an additional phase transition inside the Kondo
insulating state in two dimensions, which results from the
appearance of the topological Kondo insulator. On the other hand,
we observe that the Kondo insulating state is distinguished into
three insulating phases in three dimensions, identified with the
weak topological Kondo insulator, the strong topological Kondo
insulator, and the normal Kondo insulator, respectively, and
classified by Z$_{2}$ topological indices. We discuss the
possibility of novel quantum criticality between the
fractionalized Fermi liquid and the topological Kondo insulator,
where the band inversion occurs with the formation of the
heavy-fermion band at the same time.
\end{abstract}

\maketitle

\section{Introduction}

An effective field theory approach has been playing an important
role in predicting novel quantum states of matter. They are well
developed chiral anomaly for quantum number fractionalization,
\cite{Chiral_Anomaly} parity anomaly for quantum Hall effect,
\cite{Parity_Anomaly} and an SU(2) global anomaly
\cite{Witten_Anomaly} for topological insulators in three
dimensions \cite{Ryu_TI}. However, the field theory approach is
not enough to search such quantum matter in real materials. It is
necessary to construct or find the corresponding lattice model,
which can result in our wishful effective field theory at low
energies. In particular, the lattice model can be solved almost
exactly based on numerical simulations, helping us understand the
connection between the lattice model and effective field theory
and verifying the possibility of novel quantum phenomena.

The Su-Schrieffer-Heeger model, which describes one dimensional
electrons coupled with lattices, confirms the existence of $e/2$
fractional electric charge, carried by domain wall solitons.
\cite{Su_Schrieffer_Heeger_Model} The Haldane model, which
describes two dimensional electrons with a next nearest neighbor
complex hopping parameter on the honeycomb lattice, explains the
integer quantum Hall effect without Landau levels formed from
external magnetic field, \cite{Haldane_Model} where each Dirac
band carries a nontrivial topological quantum number (Chern
number), identified with the quantized Hall conductance.
\cite{TKNN} Recently, the Haldane model has been generalized into
the case with time reversal invariance, where the spin-orbit
coupling serves an effective magnetic flux, oppositely assigned to
spin up and down electrons, thus regarded as two copies of Haldane
models. The Kane-Mele model has proposed an interesting insulating
state, called a topological insulator, where the quantum spin Hall
effect appears when the $z$-component of spin quantum number is
preserved, but generically classified by the Z$_{2}$ topological
quantum number with spin non-conserving terms.
\cite{Kane_Mele_Model} The concept of the topological insulator
was extended into the three dimensional case,
\cite{3D_TI_Z2index_FuKane,3D_TI_Z2index_Balents,3D_TI_Z2index_Roy}
characterized by the Z$_{2}$ topological index which counts the
number of band inversions with modular two. Odd number of band
inversions cause odd number of Dirac bands, identified with
normalizable fermion zero modes localized at each two dimensional
surface, where one of them at least is protected against time
reversal invariant perturbations due to a topological origin.

Considering that these models describe non-interacting electrons
basically, an immediate and important question is on the role of
electron correlations in such topological states of matter.
\cite{Balents_MIT_SO} Actually, this direction of research has
been performed intensively, in particular, focusing on the
emergence of gapped spin liquids, \cite{Balents_Review} where
spinons with fractional spin quantum number $1/2$ appear as
elementary excitations. The quantum dimer model on the triangular
lattice has confirmed the existence of a short-ranged resonating
valence bond state, identified with Z$_{2}$ spin liquid.
\cite{QDM_Z2SL} This gapped spin liquid state has been claimed to
appear in strongly correlated electrons on the honeycomb lattice.
\cite{Dirac_Z2SL} Such a spirit has been also realized in strongly
correlated spinless fermions, \cite{Fulde_Fractional_Charge} where
charge fractionalization occurs in geometrically frustrated
lattices. Artificial spin models but exactly solvable, represented
as the Kitaev model, \cite{Kitaev_Model} have been investigated,
proving the existence of topologically non-trivial quantum states
of matter, for example, a non-abelian fractional quantum Hall
state, which allows Majorana fermions. \cite{Moore_Green} Such
exotica has been also pursued near quantum criticality, referred
as deconfined quantum criticality, where an effective field theory
approach had suggested the emergence of spin quantum number
fractionalization at an antiferromagnet to valence bond solid
quantum critical point, \cite{DQCP} and numerical simulations for
an extended Heisenberg model on the square lattice confirmed this
scenario with some modifications. \cite{Sandvick,Balents_etal}

Recently, the fractional quantum Hall effect has been argued to
appear in a certain type of lattice models, whose characteristic
feature is the existence of an almost flat band with a non-trivial
Chern number. \cite{FQHE_Lattice_Model} Electron correlations for
this partially filled flat band with the Chern number were
demonstrated to result in the fractional quantum Hall phase. An
immediate and interesting question is how to construct the lattice
model, which shows the so called fractional topological insulator.
An effective field theory approach argued the possibility of the
fractional topological insulator,
\cite{FTI_EFT_Theta1,FTI_EFT_Theta2} characterized by the fact
that the surface Dirac fermion carries a fractional electric
charge, which should be distinguished from the fractionalized
``normal" insulator, not allowing gapless surface modes. A
hard-core boson model was proposed on the diamond lattice, where
the boson is assumed to fractionalize into two fermions with $e/2$
fractional charge, forming a topological band insulator for such
fractionalized fermions and displaying the fractional
magnetoelectric effect. \cite{FTI_Lattice_Model} Introducing
electron correlations into the Kane-Mele model,
\cite{Kane_Mele_Hubbard_Model1,Kane_Mele_Hubbard_Model2} an
interesting spin liquid state was proposed to exhibit an integer
quantum spin Hall effect in the case of $S_{z}$ conservation.

In this paper we investigate an effective Anderson lattice model
for the so called topological Kondo insulator, regarded as one of
the most studied models, particulary, for non-Fermi liquid physics
near heavy fermion quantum criticality. \cite{HFQCP_Review} The
existence of the topological Kondo insulator was pointed out
recently, where an emergent ``spin"-dependent Kondo effect gives
rise to an inversion for heavy-fermion bands, which turns out to
be responsible for the topological Kondo insulator.
\cite{TKI_Coleman} Resorting to the U(1) slave-boson mean-field
analysis, we uncover an additional phase transition inside the
Kondo insulating state in two dimensions, which results from the
appearance of the topological Kondo insulator. On the other hand,
we observe that the Kondo insulating state is distinguished into
three insulating phases in three dimensions, identified with the
weak topological Kondo insulator, the strong topological Kondo
insulator, and the normal Kondo insulator, respectively, and
classified by Z$_{2}$ topological indices. Such phase transitions
inside the Kondo insulator are described by gap closing. We derive
an effective Dirac theory near the gap closing momentum point for
the phase transition from the strong topological Kondo insulator
to the normal Kondo insulator.

\section{Phase diagram for a topological Kondo insulator}

\subsection{An effective Anderson lattice model}

In order to consider the topological nature of the heavy electron
system, we construct the simplest model which introduces one
$s$-orbital and one $f$-orbital in the unit cell. Conduction
electrons are described by
\begin{eqnarray}
&& H_c = \sum_{\boldsymbol k} \sum_{\sigma}
(\varepsilon_{\boldsymbol k}^{c} - \mu_{c}) c_{{\boldsymbol
k}\sigma}^{\dagger}c_{{\boldsymbol k}\sigma},
\end{eqnarray}
where $\varepsilon_{\boldsymbol k}^{c}$ is the dispersion relation
of the conduction electron with momentum ${\boldsymbol k}$ and
real spin $\sigma$, and $\mu_{c}$ is the electron chemical
potential. The $f$-electron Hamiltonian is given by the site
energy $\epsilon_{f}-\mu_{c}$ and the on-site Coulomb interaction
$U$,
\begin{eqnarray}
&& H_f = \sum_{\boldsymbol i} \sum_{\tau} (\epsilon_{f}-\mu_{c})
f_{{\boldsymbol i}\tau}^{\dagger} f_{{\boldsymbol i}\tau} + U
\sum_{\boldsymbol i} n_{f{\boldsymbol i}\tau} n_{f{\boldsymbol
i}-\tau},
\end{eqnarray}
where $n_{f{\boldsymbol i}\tau} = f_{{\boldsymbol
i}\tau}^{\dagger} f_{{\boldsymbol i}\tau}$ is the density operator
of the $f$-electron with pseudo-spin $\tau$ belonging to one
representation $\gamma$ of $j=5/2$ multiplet at site ${\boldsymbol
i}$. $|{\boldsymbol i}-\tau\rangle$ is the time reversal partner
of $|{\boldsymbol i}\tau\rangle$ in the Kramers doublet.
Representations other than the $\gamma$ representation are assumed
to be irrelevant.

An essential point is how these itinerant and localized electrons
are coupled, generically given by the hybridization term
\begin{eqnarray}
&& H_{hyb} = \sum_{\boldsymbol k} \sum_{\sigma,\tau}
V_{\sigma\tau}({\boldsymbol k}) c_{{\boldsymbol
k}\sigma}^{\dagger}f_{{\boldsymbol k}\tau} + H.c. ,
\end{eqnarray}
where $f_{{\boldsymbol k}\tau}$ is the Fourier ${\boldsymbol
k}$-component of $f_{{\boldsymbol i}\tau}$. In order to determine
the structure of $V_{\sigma\tau}({\boldsymbol k})$, we have to
specify the representation $\gamma$ of $f$-electron and its
surroundings associated with conduction electron sites. If we
restrict the $f$-electron states onto the $j=5/2$ multiplet, the
hopping matrix is described by
\begin{eqnarray}
\langle {\boldsymbol i}s\sigma|H_{hyb}|{\boldsymbol
j}\gamma\tau\rangle &\simeq& \langle {\boldsymbol
i}s\sigma|H_{hyb}|{\boldsymbol j}\eta\sigma\rangle \langle
{\boldsymbol j}\eta\sigma|{\boldsymbol j}m\sigma\rangle \nn
&\times& \langle {\boldsymbol j}m\sigma|{\boldsymbol
j}j=\frac{5}{2}\mu\rangle \langle {\boldsymbol
j}j=\frac{5}{2}\mu|{\boldsymbol j}\gamma\tau\rangle ,
\end{eqnarray}
where $\eta$ is the representation and basis of the cubic harmonic
oscillator, $m$ is the $z$-component of the orbital angular
momentum of $f$-electron, and $\mu$ is the $z$-component of the
total angular momentum. The approximate equality comes from the
ignorance of the $j=7/2$ multiplet in estimating the matrix
element.

Resorting to the table of $\langle {\boldsymbol
i}s\sigma|H_{hyb}|{\boldsymbol j}\eta\sigma\rangle$ in Ref.
\cite{Takegahara_Overlap}, we can estimate s-f integrals as
follows
\begin{eqnarray}
  \langle {\boldsymbol i}s|H_{hyb}|{\boldsymbol j}xyz\rangle=\sqrt{15}lmn(sf\sigma),\\
  \langle {\boldsymbol i}s|H_{hyb}|{\boldsymbol j}x(5x^2-3r^2)\rangle
   =\frac{1}{2}l(5l^2-3)(sf\sigma),\\
  \langle {\boldsymbol i}s|H_{hyb}|{\boldsymbol j}x(y^2-z^2)\rangle
   =\frac{1}{2}\sqrt{15}l(m^2-n^2)(sf\sigma),
\end{eqnarray}
where $l,m,n$ are the direction cosine of the vector $\boldsymbol{
i} - \boldsymbol{j}$. The factor of $\langle {\boldsymbol
j}\eta\sigma|{\boldsymbol j}m\sigma\rangle$ corresponds to the
weight of spherical harmonics in cubic harmonics. $\langle
{\boldsymbol j}m\sigma|{\boldsymbol j}j=\frac{5}{2}\mu\rangle$ is
nothing but the Clebsch-Gordan coefficient. The last factor of
$\langle {\boldsymbol j}j=\frac{5}{2}\mu|{\boldsymbol
j}\gamma\tau\rangle$ of Eq. (4) is the weight of
$|j=5/2\mu\rangle$ in $|\gamma\tau\rangle$. Using the bases in the
cubic crystalline electric field for $|\gamma\tau\rangle$, we
obtain the table of $\langle\eta\sigma|\gamma\tau\rangle$ in Table
I.

\begin{table}[t]
\caption{ $\langle\eta\sigma|\gamma\tau\rangle$. $\eta$ describes
a cubic harmonic oscillator. The wave function of
$|\gamma\tau\rangle$ is chosen by those of $j=5/2$ in the cubic
crystal structure. }
\begin{center}
\begin{tabular}{c|cccccc}
$\eta\sigma\setminus\gamma\tau$ & $\Gamma_7+$ & $\Gamma_7-$ &
 $\Gamma_{8(1)}+$ & $\Gamma_{8(1)}-$ & $\Gamma_{8(2)}+$ & $\Gamma_{8(2)}-$\\
\hline
A$_{2u}\uparrow$ & $\sqrt{\frac{3}{7}}$i & 0 & 0 & 0 & 0 & 0\\
T$_{1u}\alpha\uparrow$ &
 0 & 0 & 0 & -$\frac{3}{2\sqrt{7}}$ & 0 & $\frac{\sqrt{3}}{2\sqrt{7}}$\\
T$_{1u}\beta\uparrow$ &
 0 & 0 & 0 & $\frac{3}{2\sqrt{7}}$i & 0 & $\frac{\sqrt{3}}{2\sqrt{7}}$i\\
T$_{1u}\gamma\uparrow$ &
 0 & 0 & 0 & 0 & -$\sqrt{\frac{3}{7}}$ & 0\\
T$_{2u}\xi\uparrow$ &
 0 & $\frac{2}{\sqrt{21}}$ & 0 & $\frac{\sqrt{5}}{2\sqrt{21}}$ & 0 &
  $\frac{\sqrt{5}}{2\sqrt{7}}$\\
T$_{2u}\eta\uparrow$ &
 0 & $\frac{2}{\sqrt{21}}$i & 0 & $\frac{\sqrt{5}}{2\sqrt{21}}$i & 0 &
  -$\frac{\sqrt{5}}{2\sqrt{7}}$i\\
T$_{2u}\zeta\uparrow$ &
 $\frac{2}{\sqrt{21}}$ & 0 & -$\frac{\sqrt{5}}{\sqrt{21}}$ & 0 & 0 & 0\\
\hline
A$_{2u}\downarrow$ & 0 & $\sqrt{\frac{3}{7}}$i & 0 & 0 & 0 & 0\\
T$_{1u}\alpha\downarrow$ &
 0 & 0 & -$\frac{3}{2\sqrt{7}}$ & 0 & $\frac{\sqrt{3}}{2\sqrt{7}}$ & 0\\
T$_{1u}\beta\downarrow$ &
 0 & 0 & -$\frac{3}{2\sqrt{7}}$i & 0 & -$\frac{\sqrt{3}}{2\sqrt{7}}$i & 0\\
T$_{1u}\gamma\downarrow$ &
 0 & 0 & 0 & 0 & 0 & $\sqrt{\frac{3}{7}}$\\
T$_{2u}\xi\downarrow$ &
 $\frac{2}{\sqrt{21}}$ & 0 & $\frac{\sqrt{5}}{2\sqrt{21}}$ & 0 &
  $\frac{\sqrt{5}}{2\sqrt{7}}$ & 0\\
T$_{2u}\eta\downarrow$ &
 -$\frac{2}{\sqrt{21}}$i & 0 & -$\frac{\sqrt{5}}{2\sqrt{21}}$i & 0 &
  $\frac{\sqrt{5}}{2\sqrt{7}}$i & 0\\
T$_{2u}\zeta\uparrow$ &
 0 & -$\frac{2}{\sqrt{21}}$ & 0 & $\frac{\sqrt{5}}{\sqrt{21}}$ & 0 & 0\\
\end{tabular}
\end{center}
\label{d4h}
\end{table}

Next, we specify the hybridization process as the orbital of the
conduction electron is situated at the same position like the case
of $6s$-electron states of rare earth ions. Then, there is no
local hybridization between $s$- and $f$-electrons. Therefore, the
main hybridization process results from the nearest neighbor
hopping from the $f$-electron state at a site ${\boldsymbol i}$ to
the $s$-electron state at a neighboring site ${\boldsymbol i+e}$,
where $e$ is a vector connecting with a neighboring unit cell.


Using the hybridization matrix discussed above, we obtain the
following matrix element for the hopping to the (100) direction,
\begin{eqnarray}
  \langle {\boldsymbol i+x}s\uparrow|H_{hyb}|{\boldsymbol i}\Gamma_{8(1)}-\rangle
& =&\langle {\boldsymbol i+x}s\downarrow|H_{hyb}|{\boldsymbol
i}\Gamma_{8(1)}+\rangle  \nn
&=&-\frac{3}{2\sqrt{7}}(sf\sigma) ,\\
  \langle {\boldsymbol i+x}s\uparrow|H_{hyb}|{\boldsymbol i}\Gamma_{8(2)}-\rangle
    &=&\langle {\boldsymbol i+x}s\downarrow|H_{hyb}|{\boldsymbol
    i}\Gamma_{8(2)}+\rangle \nn
    &=&\frac{\sqrt{3}}{2\sqrt{7}}(sf\sigma) ,
\end{eqnarray}
for the hopping matrix element to (010) direction,
\begin{eqnarray}
\langle {\boldsymbol i+y}s\uparrow|H_{hyb}|{\boldsymbol
i}\Gamma_{8(1)}-\rangle
 &=& -\langle {\boldsymbol i+y}s\downarrow|H_{hyb}|{\boldsymbol
 i}\Gamma_{8(1)}+\rangle \nn
 &=& \frac{3}{2\sqrt{7}}{\rm i}(sf\sigma),\\
 \langle {\boldsymbol i+y}s\uparrow|H_{hyb}|{\boldsymbol i}\Gamma_{8(2)}-\rangle
    &=&-\langle {\boldsymbol i+y}s\downarrow|H_{hyb}|{\boldsymbol
    i}\Gamma_{8(2)}+\rangle \nn
    &=&\frac{\sqrt{3}}{2\sqrt{7}}{\rm i}(sf\sigma),
\end{eqnarray}
for the hopping matrix element to (001) direction,
\begin{eqnarray}
\langle {\boldsymbol i+z}s\uparrow|H_{hyb}|{\boldsymbol
i}\Gamma_{8(2)}+\rangle &=&-\langle {\boldsymbol
i+z}s\downarrow|H_{hyb}|{\boldsymbol
    i}\Gamma_{8(2)}-\rangle \nn
   &=&-\sqrt{\frac{3}{7}}(sf\sigma),
\end{eqnarray}
respectively.

\begin{widetext}

Introducing $V_{sf}=\frac{\sqrt{3}}{2\sqrt{7}}(sf\sigma)$ for
simplicity, we obtain the hybridization Hamiltonian to
(100)-direction as
\begin{eqnarray}
  H_{hyb}^{(100)}&=&V_{sf}\sum_{\boldsymbol k}(
   \left[
    \begin{array}{cc}
     c_{{\boldsymbol k}\uparrow}^{\dagger} & c_{{\boldsymbol k}\downarrow}^{\dagger}\\
    \end{array}
   \right]
   \left[
    \begin{array}{cc}
     0 & 2\sqrt{3}{\rm i}\sin{k_x}\\
     2\sqrt{3}{\rm i}\sin{k_x} & 0\\
    \end{array}
   \right]
   \left[
    \begin{array}{c}
     f_{{\boldsymbol k}\Gamma_{8(1)}+}\\
     f_{{\boldsymbol k}\Gamma_{8(1)}-}\\
    \end{array}
   \right]\nonumber\\
  &&\hspace{10mm}+\left[
    \begin{array}{cc}
     c_{{\boldsymbol k}\uparrow}^{\dagger} & c_{{\boldsymbol k}\downarrow}^{\dagger}\\
    \end{array}
   \right]
   \left[
    \begin{array}{cc}
     0 & -2{\rm i}\sin{k_x}\\
     -2{\rm i}\sin{k_x} & 0\\
    \end{array}
   \right]
   \left[
    \begin{array}{c}
     f_{{\boldsymbol k}\Gamma_{8(2)}+}\\
     f_{{\boldsymbol k}\Gamma_{8(2)}-}\\
    \end{array}
   \right]
  )+H.c.,
\end{eqnarray}
for (010)-direction as
\begin{eqnarray}
  H_{hyb}^{(010)}&=&V_{sf}\sum_{\boldsymbol k}(
   \left[
    \begin{array}{cc}
     c_{{\boldsymbol k}\uparrow}^{\dagger} & c_{{\boldsymbol k}\downarrow}^{\dagger}\\
    \end{array}
   \right]
   \left[
    \begin{array}{cc}
     0 & 2\sqrt{3}\sin{k_y}\\
     -2\sqrt{3}\sin{k_y} & 0\\
    \end{array}
   \right]
   \left[
    \begin{array}{c}
     f_{{\boldsymbol k}\Gamma_{8(1)}+}\\
     f_{{\boldsymbol k}\Gamma_{8(1)}-}\\
    \end{array}
   \right]\nonumber\\
  &&\hspace{10mm}+\left[
    \begin{array}{cc}
     c_{{\boldsymbol k}\uparrow}^{\dagger} & c_{{\boldsymbol k}\downarrow}^{\dagger}\\
    \end{array}
   \right]
   \left[
    \begin{array}{cc}
     0 & 2\sin{k_y}\\
     -2\sin{k_y} & 0\\
    \end{array}
   \right]
   \left[
    \begin{array}{c}
     f_{{\boldsymbol k}\Gamma_{8(2)}+}\\
     f_{{\boldsymbol k}\Gamma_{8(2)}-}\\
    \end{array}
   \right]
  )+H.c.,
\end{eqnarray}
for (001)-direction as
\begin{eqnarray}
  H_{hyb}^{(001)}&=&V_{sf}\sum_{\boldsymbol k}
   \left[
    \begin{array}{cc}
     c_{{\boldsymbol k}\uparrow}^{\dagger} & c_{{\boldsymbol k}\downarrow}^{\dagger}\\
    \end{array}
   \right]
   \left[
    \begin{array}{cc}
     4{\rm i}\sin{k_z} & 0\\
     0 & -4{\rm i}\sin{k_z}\\
    \end{array}
   \right]
   \left[
    \begin{array}{c}
     f_{{\boldsymbol k}\Gamma_{8(2)}+}\\
     f_{{\boldsymbol k}\Gamma_{8(2)}-}\\
    \end{array}
   \right]+H.c..
\end{eqnarray}

Gathering all these terms, the resulting hybridization Hamiltonian
is given by
\begin{eqnarray}
  H_{hyb} = V_{sf} \sum_{\boldsymbol k}
   \left[
    \begin{array}{cc}
     c_{{\boldsymbol k}\uparrow}^{\dagger} & c_{{\boldsymbol k}\downarrow}^{\dagger}\\
    \end{array}
   \right]
   \left(\hat{V}_{\Gamma_{8(1)}}({\boldsymbol k})
   \left[
    \begin{array}{c}
     f_{{\boldsymbol k}\Gamma_{8(1)}+}\\
     f_{{\boldsymbol k}\Gamma_{8(1)}-}\\
    \end{array}
   \right]
   +\hat{V}_{\Gamma_{8(2)}}({\boldsymbol k})
   \left[
    \begin{array}{c}
     f_{{\boldsymbol k}\Gamma_{8(2)}+}\\
     f_{{\boldsymbol k}\Gamma_{8(2)}-}\\
    \end{array}
   \right]\right)+H.c.,
\end{eqnarray}
\end{widetext}
where the hybridization matrix
\begin{eqnarray}
  \hat{V}_{\Gamma}({\boldsymbol k})
   ={\boldsymbol d}_{\Gamma}({\boldsymbol k})\cdot\hat{\boldsymbol \sigma}
\end{eqnarray}
is
\begin{eqnarray}
  {\boldsymbol d}_{\Gamma_{8(1)}}({\boldsymbol k})
   =(2\sqrt{3}{\rm i} \sin{k_x}, 2\sqrt{3}{\rm i} \sin{k_y}, 0),\\
  {\boldsymbol d}_{\Gamma_{8(2)}}({\boldsymbol k})
   =(-2{\rm i} \sin{k_x}, 2{\rm i} \sin{k_y},
     4{\rm i} \sin{k_z}).
\end{eqnarray}

We note that this model Hamiltonian describes the nearest neighbor
hybridization process between $s$- and $f$-electrons of rare earth
ions in the simple cubic lattice. In appendix A we discuss how the
hybridization Hamiltonian is modified when the lattice symmetry is
lowered from the simple cubic to the tetragonal.

\begin{widetext}

\subsection{U(1) slave-boson mean-field analysis}

We start from an effective Anderson lattice model
\begin{eqnarray}
H &=& \sum_{\boldsymbol{k}} \sum_{\sigma}
(\varepsilon_{\boldsymbol{k}}^{c}-\mu_{c})
c_{\boldsymbol{k}\sigma}^{\dagger}c_{\boldsymbol{k}\sigma} +
\sum_{\boldsymbol{i}} \sum_{\alpha} (\epsilon_{f}-\mu_{c})
f_{\boldsymbol{i}\alpha}^{\dagger}f_{\boldsymbol{i}\alpha}   +
\epsilon \sum_{\boldsymbol{i j}} \sum_{\alpha\alpha'}
t_{\boldsymbol{ij}\alpha\alpha'}
f_{\boldsymbol{i}\alpha}^{\dagger}f_{\boldsymbol{j}\alpha'}
+\frac{U}{2} \sum_{\boldsymbol{i}} \sum_{\alpha}
f_{\boldsymbol{i}\alpha}^{\dagger}f_{\boldsymbol{i}\alpha}
f_{\boldsymbol{i}-\alpha}^{\dagger } f_{\boldsymbol{i}-\alpha}
\nonumber
\\ && + V \sum_{\boldsymbol{k}} \sum_{\boldsymbol{i}} \sum_{\alpha\sigma} [{\boldsymbol
d}_{\Gamma}({\boldsymbol k})\cdot\hat{\boldsymbol
\sigma}_{\sigma\alpha} c_{\boldsymbol{k}\sigma
}^{\dagger}f_{\boldsymbol{i}\alpha}e^{-i\boldsymbol{k}\cdot\boldsymbol{r}_{i}}+
H.c. ] ,
\end{eqnarray}
where $t_{\boldsymbol{ij}\alpha\alpha'}$ is the hopping integral
of $f$-electrons with a parameter $\epsilon \ll 1$ and the
subscript $sf$ in $V_{sf}$ is omitted for simplicity. This
effective hopping for localized fermions is introduced
phenomenologically to describe possible phase transitions from the
``fractionalized" Fermi liquid state \cite{Fractionalized_FL_KLM}
to the Kondo insulating phase at zero temperature, which can be
regarded to result from Ruderman-Kittel-Kasuya-Yoshida (RKKY) spin
correlations effectively in an intermediate energy scale.
\cite{Pepin_ALM_KB} The hybridization coupling term is given by
Eqs. (18) and (19) for the simple cubic case. \cite{Form_Factor}

We take the U(1) slave-boson representation for the Kondo effect
in the strong coupling limit, where the localized fermion is
expressed as $f_{\boldsymbol{i}\alpha}\rightarrow
b_{\boldsymbol{i}}^{\dagger}f_{\boldsymbol{i}\alpha}$ with the
single occupancy constraint
$b_{\boldsymbol{i}}^{\dagger}b_{\boldsymbol{i}}+\sum\limits_{\alpha=\pm}
f_{\boldsymbol{i}\alpha}^{\dagger}f_{\boldsymbol{i}\alpha} = 1$.
Then, the effective Lagrangian is given by
\begin{eqnarray}
&& L = \sum_{\boldsymbol{k}} \sum_{\sigma}
c_{\boldsymbol{k}\sigma}^{\dagger}(\partial_{\tau} - \mu_{c} +
\varepsilon_{\boldsymbol{k}}^{c})c_{\boldsymbol{k}\sigma} + V
\sum_{\boldsymbol{k}}\sum_{\boldsymbol{i}}\sum_{\alpha\sigma}
[\boldsymbol{\Phi}_{\alpha\sigma}(\boldsymbol{k})c_{\boldsymbol{k}\sigma}^{\dagger
}b_{\boldsymbol{i}}^{\dagger}f_{\boldsymbol{i}\alpha}e^{-i\boldsymbol{k}\cdot\boldsymbol{r}_{i}}
+ H.c. ] \nn && + \sum_{\boldsymbol{i}} \sum_{\alpha}
f_{\boldsymbol{i}\alpha}^{\dagger}(\partial_{\tau } + \epsilon_{f}
- \mu_{c} + i \lambda_{\boldsymbol{i}})f_{\boldsymbol{i}\alpha} -
\epsilon \sum_{\boldsymbol{ij}} \sum_{\alpha \alpha^{\prime}}
t_{\boldsymbol{ij}\alpha\alpha^{\prime}}f_{\boldsymbol{i}\alpha}^{\dagger}
b_{\boldsymbol{i}}^{\dagger}b_{\boldsymbol{j}}f_{\boldsymbol{j}\alpha^{\prime}}
+ i \sum_{\boldsymbol{i}} \lambda_{\boldsymbol{i}}
(b_{\boldsymbol{i}}^{\dagger}b_{\boldsymbol{i}} - 1)   ,
\end{eqnarray}
where $\lambda_{\boldsymbol{i}}$ is the Lagrange multiplier field
for the constraint and
$\boldsymbol{\Phi}_{\alpha\sigma}(\boldsymbol{k}) \equiv
{\boldsymbol d}_{\Gamma}({\boldsymbol k})\cdot\hat{\boldsymbol
\sigma}_{\sigma\alpha}$.

Taking $b_{\boldsymbol{i}} \rightarrow b \equiv \langle
b_{\boldsymbol{i}} \rangle$ with $i \lambda_{\boldsymbol{i}}
\rightarrow \lambda$ in the saddle-point approximation, we obtain
two self-consistent equations for $b$ and $\lambda$, given by
\begin{eqnarray} && \lambda =
V^{2} \sum_{\boldsymbol{k}} \sum_{\alpha\alpha^{\prime}\sigma }
\boldsymbol{\Phi}_{\alpha\sigma}(\boldsymbol{k})
\boldsymbol{\Phi}_{\alpha^{\prime}\sigma}^{\ast}(\boldsymbol{k})
\frac{1}{\pi} \int_{-\infty}^{\infty} d \omega n_{F}(\omega){\rm
Im}\dfrac{G_{\alpha\alpha^{\prime}
}^{f}(\omega+i0^{+},\boldsymbol{k})}{\omega-\varepsilon_{\boldsymbol{k}}^{c}+\mu_{c}+i0^{+}}
, \\ && b^{2}-\frac{1}{\pi} \sum_{\boldsymbol{k}} \sum_{\alpha}
\int_{-\infty}^{\infty} d \omega n_{F}(\omega) {\rm
Im}G_{\alpha\alpha}^{f}(\omega+i0^{+},\boldsymbol{k}) = 1 ,
\end{eqnarray}
respectively. $\mb{G}^{f}(\omega,\boldsymbol{k})$ is the spinon
Green's function, given by
\begin{equation}
\mb{G}^{f}(\omega,\boldsymbol{k}) = \Big[  (\omega-\epsilon_{f}
+\mu_{c}-\lambda)\mb{1}+\epsilon \mb{t}(\boldsymbol{k})
-V^{2}b^{2}\frac{\mb{\Phi}(\boldsymbol{k}) \cdot
\mb{\Phi}^{\ast}(\boldsymbol{k})}{\omega-\varepsilon_{\boldsymbol{k}}^{c}+\mu_{c}}\Big]^{-1}
,
\end{equation}
\end{widetext}
where
$[\boldsymbol{t}(\boldsymbol{k})]_{\alpha\alpha'} \equiv \sum_{ij}
[\boldsymbol{t}_{ij}]_{\alpha\alpha'} \exp[i \boldsymbol{k} \cdot
(\boldsymbol{r}_i-\boldsymbol{r}_{j})]$.
$n_{F}(\omega)=1/(\exp(\omega/T)+1)$ is the Fermi-Dirac
distribution function.


\subsection{Phase diagram}

It is natural to expect a phase transition from the fractionalized
Fermi liquid to the Kondo insulating state, described by the
emergence of the Kondo effect ($b \not= 0$) above a critical
strength of hybridization. An interesting aspect of this effective
lattice model is that there can exist additional phase transitions
inside the Kondo insulating phase, not described by the holon
condensation but characterized by the change of the Z$_{2}$
topological index.

For simplicity we take
$\boldsymbol{t}_{\alpha\alpha'}(\boldsymbol{k}) =
\delta_{\alpha\alpha'} t(\boldsymbol{k})$ with
$t(\boldsymbol{k})=\varepsilon_{\boldsymbol{k}}^{c}$. Introduction
of the spin dependence will not change possible phases but modify
critical values of $V$ associated with their phase transitions.
The dispersion relations for the heavy fermion bands are given by
\begin{eqnarray}
\lefteqn{ E_{\pm}(\boldsymbol{k})=\frac{1}{2}(
\varepsilon_{\boldsymbol{k}}^{c} - \epsilon
t(\boldsymbol{k})+\epsilon_{f}+\lambda) -\mu_{c} } \nn
&&\pm
\frac{1}{2} \sqrt {(\varepsilon_{\boldsymbol{k}}^{c} + \epsilon
t(\boldsymbol{k})-\epsilon_{f}-\lambda)^{2}
+4V^{2}b^{2}\Delta^{2}(\boldsymbol{k})},
\end{eqnarray}
respectively, where $\Delta^2(\boldsymbol{k})=\frac{1}{2} {\rm Tr}
[\mb{\Phi}(\boldsymbol{k}) \cdot
\mb{\Phi}^{\dagger}(\boldsymbol{k})]$.

In order to see how the band inversion occurs from this dispersion
relation, we consider time reversal invariant momenta which
satisfy $\boldsymbol{k}^{*}_{m} = - \boldsymbol{k}^{*}_{m} +
\mb{G}$, where $\mb{G}$ is a reciprocal lattice vector. It is
straightforward to check that there are eight time reversal
invariant momentum points in three dimensions while four in two
dimensions. The Z$_{2}$ topological index, which measures how many
times bands are twisted with modular two, has been reformulated
for such time reversal invariant momenta in the case when the
system preserves the inversion symmetry.
\cite{3D_TI_Z2index_FuKane,3D_TI_Z2index_Balents,3D_TI_Z2index_Roy}
First, we observe that the band gap in the Kondo insulating phase
closes linearly at some of the time reversal symmetry points which
satisfy $\epsilon_{f} + \lambda =
\varepsilon_{\boldsymbol{\boldsymbol{k}_{m}^{*}}}^{c} + \epsilon
t(\boldsymbol{k}_{m}^{*})$, where $\Delta(\boldsymbol{k}) \sim
|\boldsymbol{k}|$ appears near $\boldsymbol{k}_{m}^{*}$. This
identification leads us to define the parity $\delta_m = {\rm
sign}(\varepsilon_{\boldsymbol{k}_{m}^{*}}^{c} + \epsilon
t(\boldsymbol{k}_{m}^{*}) - \epsilon_{f} - \lambda)$. See appendix
B. Following Dzero {\em et al}, \cite{TKI_Coleman} we can evaluate
the Z$_{2}$ topological indices given by
\begin{eqnarray}
I_{STI} &=& \prod_{m} \delta_{m} , \\
I_{WTI}^{\alpha} &=& \prod_{m} \delta_{m}
\Big|_{(\boldsymbol{k}_{m}^{*})_{\alpha} = 0} ,
\end{eqnarray}
where $I_{STI}$ is an index for a strong topological insulator
while $I_{WTI}^{\alpha}$ are indices for a weak topological
insulator. We will see these indices changed, showing additional
phase transitions inside the Kondo insulator.

\begin{figure}[b]
\includegraphics[width=0.47\textwidth]{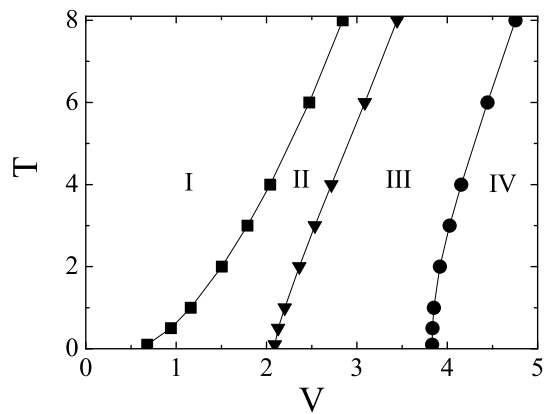}
\caption{U(1) slave-boson mean-field phase diagram for an
effective Anderson lattice model with the $\Gamma_{8(2)}$
hybridization on the simple cubic lattice. Phase I is the
fractionalized Fermi liquid state, where localized spins are
decoupled from conduction electrons ($b = 0$), forming a spinon
``Fermi" liquid state (spin liquid). The Kondo insulating phase
($b \not= 0$) is distinguished into Phase II, Phase III, and Phase
IV, classified by the topological Z$_{2}$ indices. Phase II is the
weak topological Kondo insulator and Phase III is the strong
topological Kondo insulator. Phase IV is the normal Kondo
insulator. The first transition belongs to the second order in the
saddle-point approximation, but fluctuation corrections are
possible to turn the nature of the continuous transition into the
first order. On the other hand, phase transitions inside the Kondo
insulator are also continuous, but regarded to be robust against
quantum corrections, where both hybridization and gauge
fluctuations are irrelevant. The topological aspect of the band
structure is changed by the gap closing transition. Model
parameters of $t=1$, $\epsilon=0.01$, $\epsilon_f=-6$, and
$\mu_{c}=0$ are used.} \label{fig1}
\end{figure}

We solve the mean field equations (22) and (23) with (24) for the
simple cubic lattice with
\begin{equation}
\varepsilon_{\boldsymbol{k}}^{c} = t(\boldsymbol{k})=-2t(\cos k_x
+ \cos k_{y} + \cos k_{z})
\end{equation} numerically,
and take the hybridization term of the $\Gamma_{8(2)}$ symmetry.
Our slave-boson analysis uncovers 4 different phases in three
dimensions, shown in Fig \ref{fig1}. When the hybridization
coupling is smaller than a critical value V$_{c}$, the Kondo
effect does not exist, i.e., $b = 0$ (Phase I), giving rise to
local magnetic moments decoupled from conduction electrons, where
such localized spins form a spinon ``Fermi" liquid state (spin
liquid) due to their dispersions originating from the RKKY
interaction. \cite{Fractionalized_FL_KLM} Such an exotic liquid
state may be realized in geometrically frustrated lattices or at
finite temperatures. \cite{Senthil_SL_nonzeroT} Increasing the
hybridization coupling above V$_{c}$, the Kondo effect results in
the formation of the heavy fermion band ($b \not= 0$), but the
condition of half filling for conduction electrons leads to an
insulating state instead of the heavy fermion metal. An
interesting point is that the Kondo insulating state is
distinguished into three insulating phases, classified by the
Z$_{2}$ topological indices. The phase II is characterized by the
trivial strong topological-insulator index of $I_{STI}=1$, but
nontrivial weak topological-insulator indices of
$I_{WTI}^{x}=I_{WTI}^{y}=I_{WTI}^{z}=-1$. Thus, the phase II is
identified with the weak topological Kondo insulator. The phase
III is characterized by $I_{STI}=-1$ and
$I_{WTI}^{x}=I_{WTI}^{y}=I_{WTI}^{z}=1$. As a result, it is the
strong topological Kondo insulator. The last Kondo insulating
phase, phase IV, is the conventional Kondo insulator with trivial
Z$_{2}$ indices of $I_{STI}=1$ and
$I_{WTI}^{x}=I_{WTI}^{y}=I_{WTI}^{z}=1$.

\begin{figure}[t]
\includegraphics[width=0.47\textwidth]{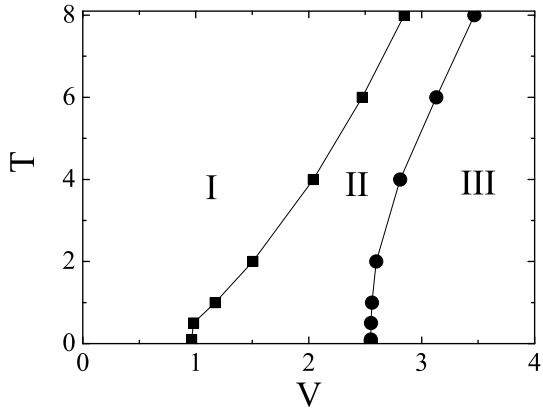}
\caption{U(1) slave-boson mean-field phase diagram for an
effective Anderson lattice model with the $\Gamma_{8(1)}$
hybridization and $\boldsymbol{k}_{z} = 0$, i.e., on the square
lattice. Phase I is the fractionalized Fermi liquid state ($b =
0$), Phase II is the topological Kondo insulator ($b \not= 0$ and
$I_{WTI}^{z} = -1$), and Phase III is the normal Kondo insulator
($b \not= 0$ and $I_{WTI}^{z} = 1$). Model parameters of $t=1$,
$\epsilon=0.01$, $\epsilon_f=-6$, and $\mu_{c}=0$ are used.}
\label{fig2}
\end{figure}

Since the first phase transition from the fractionalized Fermi
liquid to the weak topological Kondo insulator is given by the
holon condensation, i.e., the formation of the Kondo effect, it
belongs to the second order transition in the saddle-point
approximation. Two more phase transitions inside the Kondo
insulator are described by gap closing, thus also belonging to the
continuous transition. An important point is that such second
order transitions inside the Kondo insulating phase are expected
to be robust because singular corrections from fluctuations will
not exist inside the Kondo insulator. In particular, both
hybridization and gauge fluctuations will not play an important
role in these phase transitions. On the other hand, such quantum
corrections may be important for the nature of the first phase
transition from the fractionalized Fermi liquid to the weak
topological Kondo insulator, which will be discussed in the last
section.

\begin{figure}[t]
\vspace{5mm}
\caption{Top: The band structure in the Kondo insulating state
(Phase III). It is almost identical with that in the topological
Kondo insulating phase (Phase II). The electron chemical potential
lies between the band gap. Down: The band structure at the
critical point inside the Kondo insulator. Gap closing appears at
$\boldsymbol{k}_{\pi\pi}^{*} = (\pi, \pi)$, described by the Dirac
theory from Eq. (31) to Eq. (39) with $\boldsymbol{k}_{z} = 0$.
Model parameters of $t=1$, $\epsilon=0.01$, and $\epsilon_f=-6$
are used.} \label{fig3}
\end{figure}

It is straightforward to check the two dimensional case, i.e., the
effective Anderson lattice model on the square lattice. Indeed, we
performed the same slave-boson mean-field analysis in two
dimensions and found the corresponding phase diagram of Fig. 2.
The main difference from the three dimensional case is the absence
of the strong topological Kondo insulating phase, where the
quantum ``spin" Hall Kondo insulating phase exists between the
fractionalized Fermi liquid and the normal Kondo insulator. Fig. 3
display the band structures of both the topological (normal) Kondo
insulating phase and the critical point, where the gap closing
occurs at $\boldsymbol{k}_{\pi\pi}^{*} = (\pi, \pi)$.

\subsection{An effective Dirac theory for the hybridization coupling with the symmetry
$\Gamma_{8(1)}$}

It is given by the Dirac theory an effective field theory for the
phase transition from the strong topological Kondo insulator to
the normal Kondo insulator. The effective slave-boson Hamiltonian
for the topological Kondo insulator is
\begin{widetext}
\begin{eqnarray}
H &=& \sum\limits_{\boldsymbol{k}}\Psi_{\boldsymbol{k}}^{\dagger}
H(\boldsymbol{k})
\Psi_{\boldsymbol{k}}+\lambda(b^{2}-1), \\
H(\boldsymbol{k}) &=& \left(
\begin{array}[c]{cccc}
\varepsilon_{\boldsymbol{k}}^{c}-\mu_{c} & V b \Phi(\boldsymbol{k}) & 0 & 0\\
V b \Phi^{\ast}(\boldsymbol{k}) & -\epsilon t(\boldsymbol{k}
)+\widetilde{\epsilon}_{f}-\mu_{c} & 0 & 0\\
0 & 0 & \varepsilon_{\boldsymbol{k}}^{c}-\mu_{c} & -V b
\Phi^{\ast}(\boldsymbol{k})\\
0 & 0 & -V b\Phi(\boldsymbol{k}) & -\epsilon t(\boldsymbol{k}
)+\widetilde{\epsilon}_{f}-\mu_{c}
\end{array}
\right),
\end{eqnarray}
\end{widetext}
where
$\Psi_{\mathbf{k}}^{\dagger}=(c_{\mathbf{k}\uparrow}^{\dagger
},f_{\mathbf{k}-}^{\dagger},c_{\mathbf{k}\downarrow}^{\dagger},f_{\mathbf{k+}%
}^{\dagger})$ is the four component Dirac spinor with
$\widetilde{\epsilon}_{f}=\epsilon_{f}+\lambda$, and
$\Phi(\mathbf{k})=2\sqrt{3}i\sin k_{x}+2\sqrt{3}\sin k_{y}$ is the
form factor of the $\Gamma_{8(1)}$ hybridization.

Expanding $H(\boldsymbol{k})$ around the gap closing point with
time reversal symmetry,
$\boldsymbol{k}_{\pi\pi\pi}^{*}=(\pi,\pi,\pi)$, we obtain an
effective Dirac theory
\begin{eqnarray}
H(\mb{p})=\left(
\begin{array} [c]{cc}
h(\mb{p}) & 0\\
0 & h^{\ast}(-\mb{p})
\end{array}
\right)  ,
\end{eqnarray}
where
$h(\mb{p})=\varepsilon(\mb{p})\sigma_{0}+\mb{d}(\mb{p})\cdot\mb{\sigma}$
is the two by two Dirac Hamiltonian with
\begin{eqnarray}
\varepsilon(\mb{p}) &=& C-D(p_{x}^{2}+p_{y}^{2}+p_{z}^{2}), \\
\mb{d}(\mb{p}) &=& (A p_{y},A p_{x}, M(\mb{p})), \\
M(\mb{p}) &=& M- B (p_{x}^{2}+p_{y}^{2}+p_{z}^{2}), \\
C &=&
3t+\frac{3}{2}\epsilon+\frac{1}{2}\widetilde{\epsilon}_{f}-\mu_{c}, \\
D &=& \frac{1}{2}(t+\frac{\epsilon}{2}), \\
A &=& -2\sqrt{3}V b, \\
B &=& \frac{1}{2}(t-\frac{\epsilon}{2}), \\
M &=& 3t-\frac{3}{2}\epsilon-\frac{1}{2}\widetilde{\epsilon}_{f}.
\end{eqnarray}
This effective Hamiltonian turns out to be identical with that of
Ref. \cite{BHZ_Original} when $p_{x}\rightarrow p_{y}$ and
$p_{y}\rightarrow p_{x}$ are performed with $p_{z} = 0$. The
spectrum is
\begin{equation}
E_{\pm}(\mb{p})=\varepsilon(\mb{p})\pm\sqrt{A^{2}(p_{x}^{2}+p_{y}^{2})+M^{2}(\mb{p})}.
\end{equation}
Note that ${\rm sign}(M) =
\delta_{\boldsymbol{k}^{*}_{\pi\pi\pi}}$ corresponds to the parity
at the symmetry point
$\boldsymbol{k}^{*}_{\pi\pi\pi}=(\pi,\pi,\pi)$. Recall that the
Chern number is given by ${\cal C}=1$ for $M/B>0$ and ${\cal C}=0$
for $M/B<0$ in the two dimensional case. This means that the
parity $\delta_{\boldsymbol{k}^{*}_{\pi\pi\pi}}$ describes the
transition from the topological insulator to the trivial insulator
via gap closing.

In appendix C we derive an effective Dirac theory for the case of
hybridization with $\Gamma_{8(2)}$.

\section{Summary and discussion}

It is essential to construct realistic lattice models in searching
novel quantum states of matter. In this study we construct an
effective Anderson lattice model in order to study an interplay
between the topological aspect and strong correlation, regarded to
be two cornerstones for novel quantum states of matter. The
topological structure could be introduced in the spin-pseudospin
dependent hybridization between conduction electrons and localized
electrons, which originates from the interplay between the
spin-orbit interaction and crystalline electric field for
localized $f$-electrons. The spin-pseudospin dependent Kondo
effect plays essentially the same role as the spin-orbit coupling
for topological insulators, allowing the topological Kondo
insulator inside the Kondo insulating phase. Although its
existence was already pointed out in an interesting recent study
\cite{TKI_Coleman}, our detailed slave-boson analysis has
clarified its position in the phase diagram. In addition, we can
argue that the existence of the topological Kondo insulating phase
is robust against quantum corrections because hybridization
fluctuations cannot be strong inside the Kondo insulator and gauge
fluctuations are not, either.

Unfortunately, it is difficult to say that such topological states
are quite interesting because the appearance of the topological
Kondo insulator just comes from the inversion of heavy fermion
bands, basically the same as that in topological insulators of
non-interacting electrons. However, we would like to point out
that the present effective lattice model has huge potential for
novel quantum states of matter. In particular, we expect
interesting spin-ordering structures in the position of the
fractionalized Fermi liquid state, which results from the
underestimation of spin correlations in the slave-boson approach.
In this respect the saddle-point analysis based on the
slave-fermion theory \cite{Kim_SF} will open the possibility of
fruitful spin structures, where exotic ordering of localized spins
may appear as a result of the spin-dependent Kondo effect.
Recently, the integer quantum Hall effect was proposed in the
ferromagnetically Kondo coupled lattice model on geometrically
frustrated lattices, \cite{Spin_Chirality_FKLM} where the
kinetic-energy cost for conduction electrons becomes reduced due
to the emergence of an internal magnetic flux from the formation
of the spin chirality order.
%
%
%
%
In addition, it is natural to expect the possibility of novel
quantum criticality between the fractionalized Fermi liquid and
the topological Kondo insulator because the band inversion occurs
with the formation of the heavy-fermion band at the same time,
quite uncommon in the Landau-Ginzburg-Wilson description for phase
transitions. Of course, anomalous scaling near this quantum
criticality is expected to be beyond the present mean-field
analysis. Recently, two of us constructed an Eliashberg theory for
the spin density wave transition in the surface state of the three
dimensional topological insulator \cite{Kim_Takimoto_SDWQCP},
where the band reconstruction is not introduced but fluctuation
corrections are incorporated. In this study we uncovered that the
anomalous self-energy correction (the off diagonal self-energy in
the spin or pseudospin space) is essential for self-consistency.
We speculate that mathematically the same self-energy correction
via hybridization fluctuations will play an important role for
this nontrivial quantum criticality. In particular, we expect that
the characteristic feature for this quantum critical point may be
introduced in scaling of the anomalous Hall conductivity.
Numerical simulations for this model seem to be invaluable.
%
%

\section*{Acknowledgement}

This work was supported by the National Research Foundation of
Korea (NRF) grant funded by the Korea government (MEST) (No.
2011-0074542). M.-T. was also supported by the National Foundation
for Science and Technology Development (NAFOSTED) of Vietnam.

\appendix

\section{ Model Hamiltonian in tetragonal surroundings}

We consider the effect of symmetry lowering from cubic to
tetragonal. In the cubic crystal structure, the level scheme of
the $f$-electron for the $j=5/2$ multiplet is given by
\begin{eqnarray}
  D^{5/2}\downarrow O_h=\Gamma_7\oplus\Gamma_8,
\end{eqnarray}
which serves bases for the extraction of the hybridization
Hamiltonian. In the tetragonal surroundings, the $\Gamma_7(O_h)$
and $\Gamma_8(O_h)$ bases are reduced,
\begin{eqnarray}
  \Gamma_7(O_h)\downarrow D_{4h}=\Gamma_7,\\
  \Gamma_{8(1)}(O_h)\downarrow D_{4h}=\Gamma_7,\\
  \Gamma_{8(2)}(O_h)\downarrow D_{4h}=\Gamma_6 ,
\end{eqnarray}
respectively. Therefore, $\Gamma_7(O_h)$ and $\Gamma_{8(1)}(O_h)$
states can mix in the tetragonal surroundings, while
$\Gamma_{8(2)}(O_h)$ doublet splits from $\Gamma_8(O_h)$ quartet
by the tetragonal crystalline electric field.

Resorting to the above discussion, we can imagine a special case
that only one $f$-electron doublet is relevant to describe the low
energy physics in the tetragonal system. If the relevant doublet
belongs to $\Gamma_7$ in the tetragonal system, the effective
Hamiltonian will be
\begin{eqnarray}
  &&H^{\Gamma_7}=H_c+H_f+H_{hyb}^{\Gamma_7},\\
  &&H_c=\sum_{\boldsymbol k}\sum_{\sigma}
         (\varepsilon_{\boldsymbol k}^{c} - \mu_{c})
              c_{{\boldsymbol k}\sigma}^{\dagger}c_{{\boldsymbol k}\sigma},\\
  &&H_f=\sum_{\boldsymbol i}\sum_{\tau}
         (\epsilon_f - \mu_{c}) f_{{\boldsymbol i}\tau}^{\dagger}f_{{\boldsymbol i}\tau}
       +U\sum_{\boldsymbol i}n_{f{\boldsymbol i}\tau}n_{f{\boldsymbol i}-\tau},\\
  &&H_{hyb}^{\Gamma_7}=\sum_{\boldsymbol k}
   \left[
    \begin{array}{cc}
     c_{{\boldsymbol k}\uparrow}^{\dagger} & c_{{\boldsymbol k}\downarrow}^{\dagger}\\
    \end{array}
   \right]
   {\boldsymbol d}_{\Gamma_7}({\boldsymbol k})\cdot\hat{\boldsymbol \sigma}
   \left[
    \begin{array}{c}
     f_{{\boldsymbol k}+}\\
     f_{{\boldsymbol k}-}\\
    \end{array}
   \right]+H.c.,
\end{eqnarray}
with
\begin{eqnarray}
  &&\varepsilon_f=\varepsilon_{f\Gamma_7},\\
  &&{\boldsymbol d}_{\Gamma_7}({\boldsymbol k})
   =(2\sqrt{3}{\rm i}\tilde{V}_{sf}\sin{k_x},
     2\sqrt{3}{\rm i}\tilde{V}_{sf}\sin{k_y}, 0),\\
  &&f_{{\boldsymbol k}\tau}=f_{{\boldsymbol k}\Gamma_7\tau},
\end{eqnarray}
where $\epsilon_{f\Gamma_7}$ is the relevant $\Gamma_7$ doublet
corresponding to $f_{{\boldsymbol k}\Gamma_7\tau}$, and
$\tilde{V}_{sf}$ is affected from the original $V_{sf}$ by the
diagonalization in the tetragonal surroundings.

When the relevant doublet is $\Gamma_6$, the effective Hamiltonian
becomes
\begin{eqnarray}
  &&H^{\Gamma_6}=H_c+H_f+H_{hyb}^{\Gamma_6},\\
  &&H_c=\sum_{\boldsymbol k}\sum_{\sigma}
        (\varepsilon_{\boldsymbol k}^{c} - \mu_{c})
              c_{{\boldsymbol k}\sigma}^{\dagger}c_{{\boldsymbol k}\sigma},\\
  &&H_f=\sum_{\boldsymbol i}\sum_{\tau}
         (\epsilon_f - \mu_{c}) f_{{\boldsymbol i}\tau}^{\dagger}f_{{\boldsymbol i}\tau}
       +U\sum_{\boldsymbol i}n_{f{\boldsymbol i}\tau}n_{f{\boldsymbol i}-\tau},\\
  &&H_{hyb}^{\Gamma_6}=\sum_{\boldsymbol k}
   \left[
    \begin{array}{cc}
     c_{{\boldsymbol k}\uparrow}^{\dagger} & c_{{\boldsymbol k}\downarrow}^{\dagger}\\
    \end{array}
   \right]
   {\boldsymbol d}_{\Gamma_6}({\boldsymbol k})\cdot\hat{\boldsymbol \sigma}
   \left[
    \begin{array}{c}
     f_{{\boldsymbol k}+}\\
     f_{{\boldsymbol k}-}\\
    \end{array}
   \right]+h.c.,
\end{eqnarray}
with
\begin{eqnarray}
  &&\varepsilon_f=\varepsilon_{f\Gamma_6},\\
  &&{\boldsymbol d}_{\Gamma_6}({\boldsymbol k})
   =(-2{\rm i}V_{sf}\sin{k_x}, 2{\rm i}V_{sf}\sin{k_y},
     4{\rm i}V'_{sf}\sin{k_z}),\\
  &&f_{{\boldsymbol k}\tau}=f_{{\boldsymbol k}\Gamma_6\tau},
\end{eqnarray}
where $V'_{sf}$ corresponding to the hybridization to z-direction
is modified by the tetragonal surroundings. Here,
$\varepsilon_{\boldsymbol k}^{c}$ is the dispersion relation of
conduction electrons in the tetragonal surroundings.

\section{Review on the Z$_{2}$ topological index}

We review the parity eigenvalue in the Z$_{2}$ topological indices
of Eqs. (26) and (27). Generally speaking, any four by four
matrices with the hermitian property can be decomposed by the
identity matrix $\boldsymbol{I}$, five Dirac matrices
$\boldsymbol{\Gamma}^{a}$, and their ten commutators
$\boldsymbol{\Gamma}^{ab}=[\boldsymbol{\Gamma}^a,\boldsymbol{\Gamma}^b]/({\rm
2i})$. Thus, our Hamiltonian matrix $H^{\Gamma}({\boldsymbol k})$
can be expressed in terms of these 16 basis matrices. It is
convenient to choose the following representation for the Dirac
matrices,
\begin{eqnarray}
  \boldsymbol{\Gamma}^{(1,2,3,4,5)}
  =(\boldsymbol{\tau}_z\otimes\boldsymbol{\sigma}_0,\boldsymbol{\tau}_x\otimes\boldsymbol{\sigma}_0,
    \boldsymbol{\tau}_y\otimes\boldsymbol{\sigma}_x,\boldsymbol{\tau}_y\otimes\boldsymbol{\sigma}_y,
    \boldsymbol{\tau}_y\otimes\boldsymbol{\sigma}_z), \nn
\end{eqnarray}
%
%
%
%
where $\boldsymbol{\tau}_{\alpha}$ and
$\boldsymbol{\sigma}_{\alpha}$ are $2\times2$ matrices in the
orbital ($s$ and $f$) and spin spaces, respectively.

An essential aspect is that this general expansion can be
simplified near gap closing momentum points, which occur at time
reversal invariant momentum points $\boldsymbol{k}_{m}^{*}$, where
the gap closes linearly in momentum, thus the effective theory is
given by the Dirac theory. Such a Dirac theory, referred as the
Bernevig-Hughes-Zhang model, \cite{BHZ_Original} can be expanded
by only $\boldsymbol{I}$ and $\boldsymbol{\Gamma}^{a}$ as
%
%
%
%
\begin{eqnarray}
  H^{\Gamma}_{{\boldsymbol k}_m^*}(\boldsymbol{k})
  =d^{\Gamma}_0({\boldsymbol k}) \boldsymbol{I}+\sum_{a=1}^{5} \boldsymbol{d}^{\Gamma}_a({\boldsymbol k}) \boldsymbol{\Gamma}^a.
\end{eqnarray}
Among these Dirac matrices, $\boldsymbol{\Gamma}^1$ is nothing but
the parity operator $\boldsymbol{\hat{P}}$ in the present
representation,
\begin{eqnarray}
  \boldsymbol{\hat{P}}=\boldsymbol{\tau}_z\otimes\boldsymbol{\sigma}_0,
\end{eqnarray}
where $\boldsymbol{\tau}_z=+$ and $\boldsymbol{\tau}_z=-$
correspond to $s$- and $f$- orbitals, respectively. It should be
noted the following relations on parity
\begin{eqnarray}
  \boldsymbol{\hat{P}}\boldsymbol{\Gamma}^a\boldsymbol{\hat{P}}^{-1}=\boldsymbol{\Gamma}^a
   \hspace{10mm}{\rm for}\hspace{2mm}a=1,\\
  \boldsymbol{\hat{P}}\boldsymbol{\Gamma}^a\boldsymbol{\hat{P}}^{-1}=-\boldsymbol{\Gamma}^a
   \hspace{10mm}{\rm for}\hspace{2mm}a\neq 1.
\end{eqnarray}
Therefore, $\boldsymbol{\Gamma}^1$ is only parity-even Dirac
operator.

For the three dimensional system, there are eight time reversal
invariant points ${\boldsymbol k}_m^*$ in the first Brillouin
zone. We obtain two invariants at such time reversal invariant
points, given by
\begin{eqnarray}
  &&\boldsymbol{\Theta} H^{\Gamma}({\boldsymbol k}_m^*)\boldsymbol{\Theta}^{-1}=H^{\Gamma}({\boldsymbol k}_m^*),\\
  &&\boldsymbol{\hat{P}}H^{\Gamma}({\boldsymbol k}_m^*)\boldsymbol{\hat{P}}^{-1}=H^{\Gamma}({\boldsymbol
  k}_m^*) ,
\end{eqnarray}
where $\boldsymbol{\Theta}$ is the time reversal operator.
Considering the relation of Dirac matrices on parity
transformation, we can easily estimate the eigenvalue at
${\boldsymbol k}_m^*$
\begin{eqnarray}
  E^{\Gamma}({\boldsymbol k}_m^*)=d_0^{\Gamma}({\boldsymbol k}_m^*)
  \boldsymbol{I} + d_1^{\Gamma}({\boldsymbol k}_m^*) \boldsymbol{\hat{P}} ,
\end{eqnarray}
where $d_1^{\Gamma}({\boldsymbol k}_{m}^{*})$ is given by
\begin{eqnarray}
  d_1^{\Gamma}({\boldsymbol k}_{m}^{*}) = \frac{1}{2}(\varepsilon_{\boldsymbol{k}_{m}^{*}}^{c} + \epsilon
t(\boldsymbol{k}_{m}^{*}) - \epsilon_{f} - \lambda).
\end{eqnarray}
Then, the parity eigenvalue $\delta_m$ at ${\boldsymbol k}_m^*$
can expressed as follows
\begin{eqnarray}
  \delta_m=-{\rm sgn}(d_1^{\Gamma}({\boldsymbol k}_m^*))
          = - {\rm sgn}(\varepsilon_{\boldsymbol{k}_{m}^{*}}^{c} + \epsilon
t(\boldsymbol{k}_{m}^{*}) - \epsilon_{f} - \lambda). \nn
\end{eqnarray}


\section{An effective Dirac theory for the hybridization coupling with the symmetry
$\Gamma_{8(2)}$}

In the hybridization with $\Gamma_{8(2)}$ the effective
Hamiltonian is
\begin{widetext}
\begin{eqnarray}
H(\boldsymbol{k})=\left(
\begin{array}[c]{cccc}
\varepsilon_{\boldsymbol{k}}^{c} - \mu_{c} & V b(-2i\sin
k_{x}+2\sin k_{y}) &
0 & 4V b i\sin k_{z}\\
V b (2i\sin k_{x}+2\sin k_{y}) & -\epsilon t(\boldsymbol{k}
)+\widetilde{\epsilon}_{f}-\mu_{c} & 4 V b i\sin k_{z} & 0\\
0 & -4 V b i\sin k_{z} & \varepsilon_{\boldsymbol{k}}^{c} -
\mu_{c} & -V b(2i\sin k_{x}+2\sin k_{y})\\
-4 V b i\sin k_{z} & 0 & -V b(-2i\sin k_{x}+2\sin k_{y}) &
-\epsilon t(\boldsymbol{k})+\widetilde{\epsilon}_{f}-\mu_{c}
\end{array}
\right) .
\end{eqnarray}
\end{widetext}

Expanding this Hamiltonian around the
$\boldsymbol{k}_{\pi\pi\pi}^{\ast}$ point, we reach the following
expression for the Dirac theory
\begin{eqnarray}
H(\mb{p}) = \left(
\begin{array} [c]{cc}
h(\mathbf{p}) & g(\mathbf{p})\\
g^{\ast}(\mathbf{p}) & h^{\ast}(-\mathbf{p})
\end{array}
\right)  ,
\end{eqnarray}
where
\begin{eqnarray}
h(\mb{p})&=&\varepsilon(\mb{p})\sigma_{0}+\mb{d}(\mb{p})\cdot\mb{\sigma} ,\\
g(\mb{p}) &=& -2 A i p_{z}\sigma_{x} , \\
\varepsilon(\mb{p}) &=& C-D(p_{x}^{2}+p_{y}^{2}+p_{z}^{2}), \\
\mb{d}(\mb{p}) &=&(Ap_{y},-Ap_{x},M(\mb{p})), \\
M(\mb{p}) &=& M-B(p_{x}^{2}+p_{y}^{2}+p_{z}^{2}), \end{eqnarray}
\begin{eqnarray}
C &=&
3t+\frac{3}{2}\epsilon+\frac{1}{2}\widetilde{\epsilon}_{f}-\mu_{c} , \\
D &=& \frac{1}{2}(t+\frac{\epsilon}{2}), \\
A &=& 2V b, \\
B &=& \frac{1}{2}(t-\frac{\epsilon}{2}), \\
M &=& 3t-\frac{3}{2}\epsilon-\frac{1}{2}\widetilde{\epsilon}_{f}.
\end{eqnarray}


\begin{thebibliography}{99}
\bibitem{Form_Factor} The form factor in the hybridization term
can be expressed as follows
\begin{eqnarray}
[{\boldsymbol d}_{\Gamma}({\boldsymbol k})\cdot\hat{\boldsymbol
\sigma}]_{\sigma\alpha} = \sqrt{4\pi}
\sum_{j_{z}=-j}^{j}b_{j_{z}}^{\alpha}\sum_{m}a_{lm\sigma}^{j_{z}}Y_{l}^{m}(\Omega_{\boldsymbol{k}})
, \nonumber
\end{eqnarray}
where $a_{lm\sigma}^{j_{z}}$ is the Clebsh-Gordan coefficient,
$b_{j_z}$ is a coefficient specifying the crystal level, and
$Y_{l} ^{m}(\Omega_{\boldsymbol{k}})$ is the spherical harmonic
function.
\end{thebibliography}
\end{document}